\documentclass[12pt]{article}
\usepackage{a4p}
\usepackage{cite}
\usepackage{epsfig}
\usepackage{amsmath}     % math niceties

\begin{document}
%=======================================================================
%       Parameters for the title page
%=======================================================================
%
%  PPE number, Date and Version
%
\newcommand{\EPnum}     {CERN-EP/99-123}
\newcommand{\PRnum}     {OPAL Paper PR291}
\newcommand{\PNnum}     {OPAL Physics Note PN374}
\newcommand{\Date}      {September 6, 1999}
\newcommand{\Author}    {S.~Asai, A.~F\"{u}rtjes, M.~Gruw\'e, J.~Kanzaki,
S.~Komamiya, D.~Liu, G.~Mikenberg, K.~Nagai, 
H.~Neal, and R.~Van Kooten}
\newcommand{\MailAddr}  {Magali.Gruwe@cern.ch and Koichi.Nagai@cern.ch}
\newcommand{\EdBoard}   {C.~Schwick, C.~Sbarra, P.~Krieger,
and R.~Teuscher} 
\newcommand{\DraftVer}  {Version 3.0}
\newcommand{\DraftDate} {June 21st, 1999}
\newcommand{\TimeLimit} {Friday September 3rd, 14:00 CERN time} 
%=======================================================================
%       Parameters affecting the general appearance
%=======================================================================
\def\toprule{\noalign{\hrule \medskip}}
\def\midrule{\noalign{\medskip\hrule }}
\def\botrule{\noalign{\medskip\hrule }}
\setlength{\parskip}{\medskipamount}
 
%=======================================================================
%       Define symbols
%=======================================================================
\newcommand{\lumi}{182.1}
\newcommand{\ee}{{\mathrm e}^+ {\mathrm e}^-}
\newcommand{\sq}{\tilde{\mathrm q}}
\newcommand{\sg}{\tilde{\mathrm g}}
\newcommand{\su}{\tilde{\mathrm u}}
\newcommand{\sd}{\tilde{\mathrm d}}
\newcommand{\seff}{\tilde{\mathrm f}}
\newcommand{\sele}{\tilde{\mathrm e}}
\newcommand{\sell}{\tilde{\ell}}
\newcommand{\snu}{\tilde{\nu}}
\newcommand{\smu}{\tilde{\mu}}
\newcommand{\ch}{\tilde{\chi}^\pm}
\newcommand{\chp}{\tilde{\chi}^+}
\newcommand{\chm}{\tilde{\chi}^-}
\newcommand{\chpm}{\tilde{\chi}^\pm}
\newcommand{\chmp}{\tilde{\chi}^\mp}
\newcommand{\nt}{\tilde{\chi}^0}
\newcommand{\qq}{{\mathrm q}\bar{\mathrm q}}
\newcommand{\WW}{{\mathrm W}^+ {\mathrm W}^-}
\newcommand{\gv}{\gamma^*}
\newcommand{\Wenu}{{\mathrm W} {\mathrm e}\nu}
\newcommand{\ZZ}{{\mathrm Z} {\mathrm Z}}
\newcommand{\Zg}{{\mathrm Z} \gamma}
\newcommand{\Zgv}{{\mathrm Z} \gamma^*}
\newcommand{\Zvgv}{{\mathrm Z^*} \gamma^*}
\newcommand{\Wv}{{\mathrm W}^{*}}
\newcommand{\Wrv}{{\mathrm W}^{(*)}}
\newcommand{\Wvp}{{\mathrm W}^{*+}}
\newcommand{\Wvm}{{\mathrm W}^{*-}}
\newcommand{\Wvpm}{{\mathrm W}^{*\pm}}
\newcommand{\Wvrpm}{{\mathrm W}^{(*)\pm}}
\newcommand{\Z}{\mathrm Z}
\newcommand{\Zv}{{\mathrm Z}^{*}}
\newcommand{\Zrv}{{\mathrm Z}^{(*)}}
\newcommand{\ffbar}{{\mathrm f}\bar{\mathrm f}}
\newcommand{\nunu}{\nu \bar{\nu}}
\newcommand{\mumu}{\mu^+ \mu^-}
\newcommand{\tautau}{\tau^+ \tau^-}
\newcommand{\ellell}{\ell^+ \ell^-}
\newcommand{\nulqq}{\nu \ell {\mathrm q} \bar{\mathrm q}'}
\newcommand{\MZ}{m_{\mathrm Z}}
 
% Def. fuer groesser-ungefaehr:
\newcommand{\gsim}{\;\raisebox{-0.9ex}
           {$\textstyle\stackrel{\textstyle >}{\sim}$}\;}
% Def. fuer kleiner-ungefaehr:
\newcommand{\lsim}{\;\raisebox{-0.9ex}{$\textstyle\stackrel{\textstyle<}
           {\sim}$}\;}

\newcommand{\degree}    {$^\circ$}
%
%-----------------------------------------
%  Variables for machine; math mode only
%-----------------------------------------
\newcommand{\Ecm}       {E_{\mathrm{cm}}}
\newcommand{\Ebeam}     {E_{\mathrm{b}}}
\newcommand{\roots}     {\sqrt{s}}
%----------------------------------------
%  Variables for events; math mode only
%----------------------------------------
%
%     Thrust
%
\newcommand{\thrust}    {T}
\newcommand{\nthrust}   {\hat{n}_{\mathrm{thrust}}}
\newcommand{\thethr}    {\theta_{\,\mathrm{thrust}}}
\newcommand{\phithr}    {\phi_{\mathrm{thrust}}}
\newcommand{\acosthr}   {|\cos\thethr|}
\newcommand{\thejet}    {\theta_{\,\mathrm{jet}}}
\newcommand{\acosjet}   {|\cos\thejet|}
\newcommand{\thmiss}    { \theta_{\mathrm{miss}} }
\newcommand{\cosmiss}   {| \cos \thmiss |}
%
%     Energy, etc.
%
\newcommand{\Evis}      {E_{\mathrm{vis}}}
\newcommand{\Rvis}      {E_{\mathrm{vis}}\,/\roots}
\newcommand{\Mvis}      {M_{\mathrm{vis}}}
\newcommand{\Rbal}      {R_{\mathrm{bal}}}
\newcommand{\pt}        {p_{\mathrm{t}}}
%
%     Acoplanarity
%
\newcommand{\phiacop}   {\phi_{\mathrm{acop}}}
\newcommand{\axicos}{{\mid\cos\theta_a}^{\mathrm{miss}}\mid}
% \newcommand{\axicos}{\mbox{$\mid\cos\theta_a}^{\mathrm{miss}}\mid$}
%---------
%  Units
%---------
%----------------------------
%  Bibliographic references
%----------------------------
%
%     Journal names
%
\newcommand{\PhysLett}  {Phys.~Lett.}
\newcommand{\PRL} {Phys.~Rev.\ Lett.}
\newcommand{\PhysRep}   {Phys.~Rep.}
\newcommand{\PhysRev}   {Phys.~Rev.}
\newcommand{\NPhys}  {Nucl.~Phys.}
\newcommand{\NIM} {Nucl.~Instr.\ Meth.}
\newcommand{\CPC} {Comp.~Phys.\ Comm.}
\newcommand{\ZPhys}  {Z.~Phys.}
\newcommand{\IEEENS} {IEEE Trans.\ Nucl.~Sci.}
%
%     Collaboration names
%
\newcommand{\OPALColl}  {OPAL Collab.}
\newcommand{\JADEColl}  {JADE Collab.}
\newcommand{\etal}      {{\it et~al.}}
%-------
%  etc/
%-------
\newcommand{\onecol}[2] {\multicolumn{1}{#1}{#2}}
\newcommand{\ra}        {\rightarrow}   % \to can be used as well
 
%=======================================================================
%       Title Page
%=======================================================================

%-----------------------------------------------------------------------
 
\begin{titlepage}
%
%     Header
%
\begin{center}
    \large
     EUROPEAN LABORATORY FOR PARTICLE PHYSICS
\end{center}
\begin{flushright}
    \large
   \EPnum\\
%   \PRnum\\
%   \PNnum\\       %PN
   \Date
%  \DraftDate
\end{flushright}
%
%       Draft notice
%
%%\ifthenelse{\boolean{Draft}} {
%%    \begin{center}
%%      \Large\sf\bf
%%      Draft \DraftVer ,\ \ \DraftDate %\\prch1.tex
%%    \end{center}
%%   \vspace*{3mm}
%%} {}
%
%     Main title
%
\bigskip\bigskip
\begin{center}
\Large\bf\boldmath
    Search for Chargino and Neutralino Production  
  at $\sqrt{s} = 189$~GeV at LEP
\end{center}
\bigskip
\bigskip
\begin{center}{\Large The OPAL Collaboration
}\end{center}\bigskip\bigskip
\bigskip
%
%     Abstract
%
\begin{abstract}%=======================================================
A search for charginos and neutralinos, 
predicted by supersymmetric theories, is
performed using a data sample of 182.1~pb$^{-1}$ 
taken at a centre-of-mass energy of 189~GeV 
with the OPAL detector at LEP.  
No evidence for chargino or neutralino production is found.
Upper limits on chargino and neutralino pair production 
($\chp_1 \chm_1$, $\nt_1 \nt_2$) cross-sections are obtained
as a function of the chargino mass ($m_{\chpm_1}$),
the lightest neutralino mass ($m_{\nt_1}$) 
and the second lightest neutralino mass ($m_{\nt_2}$)\@.  
Within the Constrained Minimal Supersymmetric Standard Model
framework, 
and for $m_{\chpm_1} - m_{\nt_1} \geq 5$~GeV,
the 95\% confidence level lower limits on $m_{\chpm_1}$ 
are 93.6~GeV for $\tan \beta = 1.5$ and 94.1~GeV for $\tan \beta = 35$\@.
These limits are obtained assuming a universal scalar
mass $m_0 \geq$ 500~GeV\@. 
The corresponding limits for all $m_0$ are 78.0
and 71.7~GeV.
The 95\% confidence level lower limits on the lightest neutralino mass,
valid for any value of $\tan \beta$ 
are 32.8~GeV for $m_0 \geq 500$~GeV and 31.6~GeV for all $m_0$.
\end{abstract}%=========================================================

\bigskip
\bigskip
% \begin{center}
% {\large \bf Comments to {\MailAddr}\\  by {\TimeLimit}}
% {\large \bf First Draft}
% {\large \bf \DraftVer (\DraftDate)}
% \end{center}
\bigskip

\begin{center}
{\bf \large (Submitted to Phys. Lett. B)}\\
\end{center}
\end{titlepage}
%
%     Author names
%
\begin{center}{\Large        The OPAL Collaboration
}\end{center}\bigskip
\begin{center}{
%begin authorlist PLEASE DO NOT DELETE THIS COMMENT
G.\thinspace Abbiendi$^{  2}$,
K.\thinspace Ackerstaff$^{  8}$,
G.\thinspace Alexander$^{ 23}$,
J.\thinspace Allison$^{ 16}$,
K.J.\thinspace Anderson$^{  9}$,
S.\thinspace Anderson$^{ 12}$,
S.\thinspace Arcelli$^{ 17}$,
S.\thinspace Asai$^{ 24}$,
S.F.\thinspace Ashby$^{  1}$,
D.\thinspace Axen$^{ 29}$
% ...
%end authorlist PLEASE DO NOT DELETE THIS COMMENT
}\end{center}\bigskip
\bigskip
%begin institutes
$^{  1}$School of Physics and Astronomy, University of Birmingham,
Birmingham B15 2TT, UK
\newline
$^{  2}$Dipartimento di Fisica dell' Universit\`a di Bologna and INFN,
I-40126 Bologna, Italy
\newline
% ...
$^{  8}$CERN, European Organisation for Particle Physics,
CH-1211 Geneva 23, Switzerland
\newline
$^{  9}$Enrico Fermi Institute and Department of Physics,
University of Chicago, Chicago IL 60637, USA
\newline
% ...
$^{ 12}$Indiana University, Department of Physics,
Swain Hall West 117, Bloomington IN 47405, USA
\newline
% ...
$^{ 16}$Department of Physics, Schuster Laboratory, The University,
Manchester M13 9PL, UK
\newline
$^{ 17}$Department of Physics, University of Maryland,
College Park, MD 20742, USA
\newline
% ...
$^{ 23}$Department of Physics and Astronomy, Tel Aviv University,
Tel Aviv 69978, Israel
\newline
$^{ 24}$International Centre for Elementary Particle Physics and
Department of Physics, University of Tokyo, Tokyo 113-0033, and
Kobe University, Kobe 657-8501, Japan
\newline
% ...
$^{ 29}$University of British Columbia, Department of Physics,
Vancouver BC V6T 1Z1, Canada
\newline
% ...
 Debrecen, Hungary.
\newline
%end notes
\newpage
%=======================================================================
\section{Introduction}
%=======================================================================

Supersymmetric (SUSY) extensions of the Standard Model
predict the existence of charginos and neutralinos~\cite{mssm}\@.
Charginos, $\chpm_{i}$, are the mass eigenstates formed by the mixing of
the fields of the fermionic partners of the charged gauge bosons (winos)
and those of the charged Higgs bosons (charged higgsinos).
Fermionic partners of the $\gamma$, the Z boson,
and the neutral Higgs bosons
mix to form the mass eigenstates called neutralinos,
$\nt_{j}$~\footnote{In each case, the index $i={1,2}$
or $j={1~\mathrm{to}~4}$ is ordered by increasing mass.}\@.
If charginos exist and are sufficiently light,
they are pair-produced at LEP through $\gamma$- or Z-exchange 
in the $s$-channel.
For the wino component, there is an additional production process 
through scalar electron-neutrino ($\snu_{\mathrm{e}}$) 
exchange in the $t$-channel.
The production cross-section is large (several pb)
unless the scalar neutrino (sneutrino) is light, in which case
the cross-section is reduced by destructive interference
between the $s$-channel and $t$-channel diagrams~\cite{chargino}.
% The details of chargino decay depend on
% the parameters of the mixing
% and the masses of the scalar partners of the ordinary fermions.
In much of the parameter space, $\chp_1$ decays dominantly 
into $\nt_1 \ell^+ \nu$ or $\nt_1 \mathrm{q} \mathrm{\overline{q}'}$
via a virtual W boson.
For small scalar lepton masses, decays to leptons via a scalar lepton 
become important. 
R-parity conservation is assumed throughout this note.
With this assumption, the $\nt_1$ is stable and invisible
\footnote{The lightest supersymmetric particle (LSP) is either 
$\nt_1$ or the scalar neutrino.
We assume $\nt_1$ is the LSP in the direct searches described in this paper.
If the scalar neutrino is lighter than the chargino, 
$\chp_1 \rightarrow \snu \ell^+$ becomes the dominant decay mode.
For this case, we use the results of Ref.~\cite{LEP18-slepton}
to calculate limits, as mentioned in Section~5.2.}
and the experimental signature for $\chp_1 \chm_1$ events
is large missing momentum transverse
to the beam axis.

Neutralino pairs ($\nt_1 \nt_2$) can be produced through a
virtual Z or $\gamma$ ($s$-channel) or by a scalar electron ($t$-channel)
exchange~\cite{neutralino}.
The $\nt_2$ will decay into 
$\nt_1 \nunu$, $\nt_1 \ellell$ or
$\nt_1 {\mathrm q}\bar{\mathrm q}$
through a virtual Z boson,
sneutrino, slepton, squark or
a neutral SUSY Higgs boson ($\mathrm{h}^0$ or $\mathrm{A}^0$)\@.
The decay via a virtual Z is the dominant mode in most of 
the parameter space.
For small scalar lepton masses, decays to a lepton pair via a scalar lepton 
are important.
The experimental signature of $\nt_1 \nt_2$ events is
an acoplanar pair of leptons or jets. If the
mass difference between $\nt_2$ and $\nt_1$ is small,
the experimental signature becomes monojet-like.

Motivated by Grand Unification and to simplify
the physics interpretation,
the Constrained Minimal Supersymmetric
Standard Model (Constrained MSSM)~\cite{mssm, mssm2}
is used to guide the analysis but the results are also applied to more
general models.
At the grand unified (GUT) mass scale,  
in the Constrained MSSM all the gauginos 
are assumed to have a common mass, $m_{1/2}$, 
and all the sfermions
have a common mass, $m_0$.   
Details are given in Section~5.2\@.

Previous searches for       
charginos and neutralinos have been performed
using data collected near the Z peak (LEP1), 
at centre-of-mass energies ($\roots$) of 130--136~GeV~\cite{LEP15}, 
161~GeV~\cite{LEP16}, 170--172~GeV~\cite{LEP17-opal,LEP17}, 
and 183~GeV~\cite{LEP18-opal,LEP18}, with luminosities of about 
100, 5, 10, 10 and 60~pb$^{-1}$ respectively. 
No evidence for signal has been found.
% A 95\% C.L. lower limit on the $\chpm_1$ mass 
% was set at 69~GeV  
% within the CMSSM framework for
% $m_{\chpm_1} - m_{\nt_1} \geq 5$~GeV\@.

In 1998 the LEP $\ee$ collider at CERN was operated 
at $\roots$= 188.6~GeV. 
This paper reports on direct searches for charginos and neutralinos performed 
using the data sample collected at this centre-of-mass 
energy. The total integrated luminosity collected with
the OPAL detector at this energy is 182.1~pb$^{-1}$\@.
The selection criteria are similar to those used in~\cite{LEP18-opal},
but have been modified to improve the sensitivity of the analysis at the
current energy.
The description of the OPAL detector and its performance
can be found in Ref.~\cite{OPAL-detector} and~\cite{MIPPLUG}.

%=======================================================================
\section{Event Simulation}
%=======================================================================
%
Chargino and neutralino signal events are generated with
the DFGT generator~\cite{DFGT}\@ which includes spin correlations and
allows for a proper treatment of both the W boson and the Z boson 
width effects in the chargino and heavy neutralino decays.
The generator includes initial-state radiation and uses
the JETSET~7.4 package~\cite{PYTHIA} for the hadronisation
of the quark-antiquark
system in the hadronic decays of charginos and neutralinos.
SUSYGEN~\cite{SUSYGEN} is used to calculate 
the branching fractions for the Constrained 
MSSM interpretation of the analysis. 

The sources of background to the chargino and neutralino signals
are two-photon, lepton pair, multihadronic and four-fermion
processes.
Two-photon processes are the most important background
for the case of a small mass difference between the $\chpm_1$ and the $\nt_1$, 
or between the $\nt_2$ and the $\nt_1$,
since such events have small visible energy and 
small transverse momentum.
Using the Monte Carlo generators 
PHOJET~\cite{PHOJET}, PYTHIA~\cite{PYTHIA} and HERWIG~\cite{HERWIG},
hadronic events from two-photon processes are simulated
in which the invariant mass of the photon-photon
system is larger than 5.0~GeV\@.
Monte Carlo samples for four-lepton events ($\ee \ee$, $\ee \mumu$ and 
$\ee \tautau$) are generated with the Vermaseren 
program~\cite{Vermaseren}\@.  
All other four-fermion processes except for regions of phase-space
covered by the two-photon simulations, are
simulated using the grc4f generator~\cite{grc4f},
which takes into account all interferences.   
The dominant contributions come from
$\WW$, $\Wenu$, $\gamma^* \Z^{(*)}$ and $\Z\Z^{(*)}$ processes.
Lepton pairs are generated using
the KORALZ~\cite{KORALZ} generator for
$\tau^+ \tau^- (\gamma)$ and $\mumu (\gamma)$ events,
and the BHWIDE~\cite{BHWIDE} program
for $\ee \ra \ee (\gamma) $ events.
Multihadronic ($\qq (\gamma)$) events are simulated using
PYTHIA~\cite{PYTHIA}. 

Generated signal and background events are processed
through the full simulation of the OPAL detector~\cite{GOPAL}
and the same event analysis chain is applied to the simulated events
as to the data.

%=======================================================================
\section{Analysis}
%=======================================================================

Calculations of experimental variables are performed as in~\cite{LEP17-opal}.
The following preselections are applied 
to reduce background due to two-photon events
and interactions of beam particles with the beam pipe or residual gas:
(1) the number of charged tracks is required to be at least two;
(2) the observed transverse momentum of the whole event 
is required to be larger than 1.8~GeV;
(3) the energy deposited in each silicon-tungsten
forward calorimeter and in each forward detector has to be less than
2~GeV
(these detectors are located in the forward region, with polar 
angle~\footnote{A right-handed 
coordinate system is adopted,
where the $x$-axis points to the centre of the LEP ring,
and positive $z$ is along  the electron beam direction.
The angles $\theta$ and $\phi$ are the polar and azimuthal angles,
respectively.}
$|\cos \theta|>0.99$, surrounding the beam pipe);
(4) the visible invariant mass of the event has to exceed 2 GeV;
(5) there should be no signal in the MIP plug
scintillators~\footnote{The MIP plug 
scintillators~\cite{MIPPLUG}
are an array of
thin scintillating tiles with embedded wavelength shifting fibre
readout which have been installed to improve the hermiticity of 
the detector. They cover the polar angular range between 43 and
200~mrad.}\@.

%=======================================================================
\subsection{Detection of charginos}
%=======================================================================

The event sample is divided into three mutually exclusive categories, 
motivated by the topologies expected for chargino events. 
Separate analyses are applied to the preselected events in 
each category:
\begin{itemize}
\item[(A)] 
$N_{\mathrm{ch}} > 4$ and no isolated leptons,
where $N_{\mathrm{ch}}$ is the 
number of charged tracks. 
The isolated lepton selection criteria
are the same as those described in Ref.~\cite{LEP18-opal}. 
When both $\chp_1$ and $\chm_1$ decay
hadronically, signal events tend to fall 
into this category for all but the smallest values of 
$\Delta M_+ (\equiv m_{\chp_1} - m_{\nt_1})$\@.
\item[(B)] 
$N_{\mathrm{ch}} > 4$ and at least one isolated lepton.
If just one of the $\chpm_1$ decays leptonically,
signal events tend to fall 
into this category.
\item[(C)] 
$N_{\mathrm{ch}} \leq 4$.
Events tend to fall into this category if $\Delta M_+$ is small or if
both charginos decay leptonically.
\end{itemize}

The fraction of $\chp_1 \chm_1$ events falling into
category (A) is about 35-50\% for most of the $\Delta M_+$ range.
This fraction drops to less than 15\% if $\Delta M_+$ is smaller than 5~GeV,
since the average charged track multiplicity of the events becomes small.
Similarly, 
the fraction of events falling into category (B) is also about 35-50\% 
for most of the $\Delta M_+$ range and is less than 10\% if 
$\Delta M_+$ is smaller than 5~GeV.
In contrast, when $\Delta M_+$ is smaller than 10~GeV,
the fraction of events falling into category (C) is greater than about
50\%\@, while if $\Delta M_+$ is larger than 20~GeV, 
this fraction is about 10\%\@.

Since the chargino event topology mainly depends on the 
difference between the chargino mass and the lightest neutralino
mass, different selection criteria are applied to four 
$\Delta M_+$ regions:  
(I) $\Delta M_+ \leq 10$~GeV, 
(II) 10~GeV $< \Delta M_+ \leq m_{\chp_1}/2$,
(III) $ m_{\chp_1}/2 < \Delta M_+ \leq m_{\chp_1}-20$~GeV, and 
(IV) $ m_{\chp_1}-20$~GeV$ < \Delta M_+ \leq m_{\chp_1}$\@. 
In region (I), background events come mainly 
from the two-photon processes.
In regions (II) and (III), the main background processes are 
the four-fermion processes ($\WW$, single W and $\gamma^* \Z^{(*)}$)\@.
In these two regions the background level is modest.
In region (IV) the $\WW$ background becomes large and dominant.
Since the $\WW$ background is significant in the region of 
$\Delta M_+ > 85$~GeV where the chargino decays via an on-mass-shell
W-boson, a special analysis is applied to improve the sensitivity to
the chargino signal for $m_{\chp_1} > 85$~GeV and $\Delta M_+ \gsim 85$~GeV.
Overlap between this analysis and the region (IV) standard analysis 
is avoided by selecting the analysis which minimises the expected
cross-section limit calculated with only the expected 
number of background events.

For each region a single set of cut values is determined which
minimises the expected limit on the signal cross-section at 95\%
confidence level (C.L.) using the method of 
Ref.~\cite{PDG96}. In this procedure, only  
the expected number of background events is 
taken into account and therefore the choice of cuts is 
independent of the number of candidates actually observed.

The variables used in the selection criteria and their cut values are 
optimised in each $\Delta M_+$ region. They are identical to the ones
of Ref.~\cite{LEP18-opal} unless otherwise stated, and are therefore
only briefly described in the following sections.

%=======================================================================
\subsubsection{Analysis (A) (\boldmath$N_{\mathrm{ch}} > 4$
without isolated leptons)}
%=======================================================================
After the preselection, cuts on $E_{\mathrm{fwd}}/E_{\mathrm{vis}}$, 
$|\cos \theta_{\mathrm{miss}} |$, $|P_{\mathrm{z}}|$ and $|P_{\mathrm{z}}|/E_{\mathrm{vis}}$, 
are applied to further reduce background events
from the two-photon and multihadronic processes. 
$E_{\mathrm{vis}}$ is the total visible energy of the event, 
$E_{\mathrm{fwd}}$ is the visible energy in the region of 
$| \cos \theta |>0.9$, $\theta_{\mathrm{miss}}$ is the polar angle
of the missing momentum and $P_{\mathrm{z}}$ is the visible momentum
along the beam axis. In regions (III) and (IV), the cut values for 
$E_{\mathrm{fwd}}/E_{\mathrm{vis}}$ and $|\cos \theta_{\mathrm{miss}}
|$ have been changed with respect to Ref.~\cite{LEP18-opal}, requiring 
them to be smaller than 0.15 and 0.90 respectively.
Most of the two-photon background events are rejected 
by cuts against small $P_{\mathrm t}^{\mathrm{HCAL}}$ and $P_{\mathrm t}$,
the transverse momentum of the event measured with and without using
the hadron calorimeter, respectively.
In region (I), a cut is also applied against large 
$P_{\mathrm t}^{\mathrm{HCAL}}$ ($P_{\mathrm t}^{\mathrm{HCAL}} \le
30$~GeV) to reduce the $\WW$ background.
 
Jets are reconstructed 
using the Durham algorithm~\cite{DURHAM}
with jet resolution parameter $y_{\mathrm{cut}} = 0.005$\@. By
requiring the number of jets ($N_{\mathrm{jet}}$) to be between 3
and 5 inclusive, monojet events from 
the process $\gamma^{*} {\mathrm Z}^{(*)} \ra \qq \nunu$ are rejected 
for regions (I) and (II), and background events from 
$\qq (\gamma)$ and $\Wenu$ processes are reduced in regions (III) and (IV).

In order to determine the acoplanarity
angle~\footnote{The acoplanarity 
angle, $\phi_{\mathrm{acop}}$, is defined as 180\degree - $\phi$, where
$\phi$ is the opening angle in the ($r$,$\phi$) plane 
between the two jets.} ($\phi_{\mathrm{acop}}$), 
events passing the selection described above are forced into
two jets, again using the Durham algorithm.
Each jet is required to be far from the beam axis by cutting on its
polar angle $\theta_{\mathrm j}$ (j=1,2). 
This cut ensures a good measurement of $\phi_{\mathrm{acop}}$
and further reduces the background from $\qq (\gamma)$ and
two-photon processes.  
The acoplanarity angle is required to be larger than 15\degree\@  
to reduce the $\qq (\gamma)$ background. 
The acoplanarity angle distribution for region (III) 
is shown in Figure~1(a) before this cut.

The cut on the visible mass, $M_{\mathrm{vis}}$, is optimised
for each $\Delta M_+$ region.  
If a lepton candidate ($\ell'$) is found with an algorithm based on the
``looser'' isolation condition described in Ref.~\cite{LEP17-opal},
the energy of this lepton, $E_{\ell'}$, and 
the invariant mass calculated without this lepton,
$M_{\mathrm{had'}}$, must be different from the values expected for 
$\WW \ra \ell \nu \qq'$ events. The cuts on  $E_{\ell'}$ and 
$M_{\mathrm{had'}}$ have been optimized with respect to
Ref.~\cite{LEP18-opal}: for region (I) no cut is applied; for
region (III) $M_{\mathrm{had'}}$ is required to be smaller than
65~GeV and $E_{\ell'}$ smaller than 25~GeV.

The background from $\WW$ events and single-W events 
is efficiently suppressed by requiring that 
the highest and second highest jet energies, $E_1$ and $E_2$, be
smaller than the typical jet energy expected for the $\WW \ra$
4-jet process.
The cuts on $E_1$ and $E_2$ are identical to the ones in
Ref.~\cite{LEP18-opal} apart from region (II) where $E_1$ is required to be
between 2 and 30~GeV. In addition, in regions (III) and
(IV), if $E_1$ is larger than 40~GeV, $M_{\mathrm{vis}}$ is required
to be either smaller than 70~GeV or larger than 95~GeV.
In regions (III) and (IV) three-jet events with 
$|P_{\mathrm{z}}| < 10$~GeV are also rejected
if $\Mvis$ is close to the W mass.
These cuts reduce the $\WW \ra \tau \nu \qq'$ background with
low-energy decay products of the $\tau$.

A special analysis is applied 
in the region of $\Delta M_+ \gsim 85$~GeV, since
the event topology of the signal is very similar to 
that of $\WW \ra$~4~jets.
% This similarity is due to the small missing momentum 
% taken by the low mass neutralinos.
After selecting well contained events 
with the cuts $|\cos \theta_{\mathrm{miss}}|<0.95$,
$E_{\mathrm{fwd}}/E_{\mathrm{vis}}<0.15$ and $|P_{\mathrm{z}}|<30$~GeV,
multi-jet events with large visible energy are selected with
$N_{\mathrm{jet}} \geq 4$ and $110< E_{\mathrm{vis}} < 170$~GeV\@.
To select a clear 4-jet topology $y_{34} \geq 0.0075$,
$y_{23} \geq 0.04$ and $y_{45} \leq 0.0015$ are also required, where
$y_{\{n\}\{n+1\}}$ is defined as the minimum $y_{\mathrm{cut}}$ value
at which the reconstruction of the event switches from $n+1$ to $n$ jets. 
Events having a ``jet'' consisting of a single $\gamma$
with energy greater than 20~GeV are considered 
to be $\gamma \qq g$ and are rejected.

The numbers of observed events and background events expected from
the four different sources, for 
each $\Delta M_+$ region and for the special analysis, 
are given in Table~\ref{tab:cutAch}.
Typical detection efficiencies for $\chp_1 \chm_1$ events
are 20--65\% for $\Delta M_+$=10--80~GeV in the standard analysis, and
20--30\% for  $m_{\chp_1} \geq 85$~GeV and $\Delta M_+ \geq 85$~GeV.

\begin{table}[htb]%---------------------------------
  \centering
  \begin{tabular}{|l||c|c|c|c|c|}
    \hline%-----------------------------------------
    Region  &  I  &  II  &  III & IV  & special \\
 &
$_{\Delta M_+ \leq 10~{\mathrm{GeV}}}$  &
$_{10~{\mathrm{GeV}} < \Delta M_+}$  &
$_{m_{\chp_1}/2 < \Delta M_+}$  &
$_{m_{\chp_1}-20~{\mathrm{GeV}}}$  & \\
 & & 
$_{\leq m_{\chp_1}/2}$ & 
$_{\leq m_{\chp_1}-20~{\mathrm{GeV}}}$ &
$_{< \Delta M_+ \leq m_{\chp_1}}$ & 
$_{\Delta M_+ \geq 85~{\mathrm{GeV}}}$ \\
    \hline%-----------------------------------------
    \hline%-----------------------------------------
    background & \multicolumn{5}{|c|}{ }    \\
    \hline% ----------------------------------------
    $\gamma \gamma$    & 0.93 & 0.46 & 0.46 & 0.46 & 0.00 \\
    \hline% ----------------------------------------
    $\ellell (\gamma)$ & 0.00 & 0.00 & 0.02 & 0.02 & 0.00 \\
    \hline% ----------------------------------------
    $\qq (\gamma)$     & 0.07 & 0.22 & 0.83 & 1.83 & 4.57 \\
    \hline% ----------------------------------------
    4f               & 0.55 & 1.27 & 16.54 & 24.66 & 23.09 \\
    \hline% ----------------------------------------
    \hline% ----------------------------------------
    total bkg 
    & $ 1.5\pm0.7 $ & $2.0\pm0.5$ & $17.8\pm0.9$ & $27.0\pm1.1$ & 
$27.7\pm1.0$ \\
    \hline% ----------------------------------------
    \hline% ----------------------------------------
    observed & 2 & 3 & 22 & 29 & 31 \\
    \hline% ---------------------------------------
\end{tabular}
\caption[]{%\sl
\protect{\parbox[t]{15cm}{
The numbers of expected background events 
for various Standard Model processes, 
normalised to the integrated luminosity, 
and the number of observed 
events in each $\Delta M_+$ region and for the special analysis
for category (A).
The errors in the total background include only the statistical error
of the Monte Carlo samples.
}} }
  \label{tab:cutAch}
\end{table}%------------------------------------------------------------

%=======================================================================
\subsubsection{Analysis (B) (\boldmath$N_{\mathrm{ch}} > 4$ 
with isolated leptons)}
%=======================================================================

After the preselection, cuts on $|\cos \theta_{\mathrm{miss}}|$ and 
$E_{\mathrm{fwd}}/E_{\mathrm{vis}}$ are applied to 
reject two-photon background events. The cut on
$E_{\mathrm{fwd}}/E_{\mathrm{vis}}$ has been tightened with respect to
Ref.~\cite{LEP18-opal}: it is required to be smaller than 0.15 in
regions (I) and (II) and smaller than 0.2 in regions (III) and (IV).

In order to reject the $\WW \ra \ell \nu \qq'$ background, 
the following cuts are applied:
the momentum of isolated lepton candidates should be smaller than that expected
from decays of the W (smaller than 15~GeV, 30~GeV, 40~GeV and between 5
and 40~GeV for regions (I) to (IV) respectively) and
the invariant mass ($M_{\mathrm{had}}$) of the event calculated excluding the 
highest momentum isolated lepton
is required to be smaller than the W mass (smaller than 20~GeV and
40~GeV for regions (I) and (II), between 10 and 60~GeV for region
(III), and between
15 and 70~GeV for region (IV)).
The distribution of $M_{\mathrm{had}}$ 
after the $\phi_{\mathrm{acop}}$ cut is shown in Figure~1(b) for region (III).
As is evident in this figure, most of the $\WW$ background events are
rejected by this cut.
The invariant mass of the system formed by the missing
momentum and the most energetic isolated lepton, 
$M_{\ell {\rm miss}}$, 
is required to be larger than 110~GeV for region (IV)\@.
Finally, a $M_{\mathrm{vis}}$ cut is applied
to reject $\Wenu$ events in which 
a fake lepton candidate is found in the W$\ra \qq'(g)$ decay: it is required to
be smaller than 30~GeV for region (I),
between 25 and 85~GeV for region (III), and no requirement on 
$M_{\mathrm{vis}}$ is applied for region (IV).

A special analysis is applied 
in the region of $\Delta M_+ \gsim 85$~GeV where there is a 
large $\WW$ background.  
The selection criteria are identical to those in region (IV) 
up to the $\phi_{\mathrm{acop}}$ cut.
To reject further 
$\WW \ra \ell \nu \qq'$ events while keeping a good 
signal efficiency, 
$M_{\mathrm{had}}$ is required to be between 70 and 95~GeV,
$M_{\mathrm{vis}}$ between 90 and 125~GeV, and 
$M_{\ell {\rm miss}}$ between 90 and 130~GeV.

The numbers of observed events and background events expected from
the four different sources, for 
each $\Delta M_+$ region and for the special analysis, 
are given in Table~\ref{tab:cutBch}.
Typical detection efficiencies for  
$\chp_1 \chm_1$ events are 40--65\% for $\Delta M_+$=10--80~GeV for the 
standard analysis, and 20--30\% for  
$m_{\chp_1} \geq 85$~GeV and $\Delta M_+ \geq 85$~GeV.

\begin{table}[h]%------------------------------------------------------
\centering
\begin{tabular}{|l||c|c|c|c|c| }
\hline%-----------------------------------------------------------------
    Region  &  I  &  II  &  III & IV  & special \\
 &
$_{\Delta M_+ \leq 10~{\mathrm{GeV}}}$  &
$_{10~{\mathrm{GeV}} < \Delta M_+}$  &
$_{m_{\chp_1}/2 < \Delta M_+}$  &
$_{m_{\chp_1}-20~{\mathrm{GeV}}}$  & \\
 & & 
$_{\leq m_{\chp_1}/2}$ & 
$_{\leq m_{\chp_1}-20~{\mathrm{GeV}}}$ &
$_{< \Delta M_+ \leq m_{\chp_1}}$ & 
$_{\Delta M_+ \geq 85~{\mathrm{GeV}}}$ \\
\hline%-----------------------------------------------------------------
\hline%-----------------------------------------
background &  \multicolumn{5}{|c|}{ }   \\  
\hline% ----------------------------------------
$\gamma \gamma$    & 3.80 & 0.46 & 0.00 & 0.00 & 0.00 \\ 
\hline% ----------------------------------------
$\ellell (\gamma)$ & 0.02 & 0.04 & 0.11 & 0.05 & 0.02 \\  
\hline% ----------------------------------------
$\qq (\gamma)$     & 0.00 & 0.11 & 0.14 & 0.11 & 0.04 \\  
\hline% ----------------------------------------
4f                 & 0.62 & 1.38 & 4.30 & 6.37 & 23.45\\  
\hline% ----------------------------------------
\hline% ----------------------------------------
total bkg       
& 4.4$\pm$1.3 & 2.0$\pm$0.5 & 4.6$\pm$0.4 & 6.5$\pm$0.5 & 
23.5$\pm$0.9 \\
\hline% ----------------------------------------
\hline% ----------------------------------------
observed       & 2  &  1  &  4  &  7 & 27 \\  
\hline% ----------------------------------------
\end{tabular}
\caption[]{%\sl
  \protect{\parbox[t]{15cm}{
The numbers of expected background events 
for various Standard Model processes, 
normalised to the integrated luminosity, 
and the number of observed 
events in each $\Delta M_+$ region and for the special analysis 
for category (B).
The errors in the total background include only the statistical error
of the Monte Carlo samples.
}} 
}
\label{tab:cutBch}
\end{table}%------------------------------------------------------------

%=======================================================================
\subsubsection{Analysis (C) (\boldmath$N_{\mathrm{ch}} \leq 4$)}
%=======================================================================

This analysis is especially important for the region of 
$\Delta M_+ \leq 5$~GeV. 
Because the background varies significantly with $\Delta M_+$ in region (I), 
this region is split into 2 sub-regions (a,b).

In order to reject events 
with charged particles which escape detection in the main detector, 
the net charge of the event is required to be zero.
Since the signal is expected to have a two-lepton or two-jet topology,
events are forced into two jets using the Durham jet algorithm~\cite{DURHAM}\@.
To improve the jet assignment, each jet
must contain at least one charged track ($N_{\mathrm{ch,j}} \ge 1$), 
must have significant energy ($E_{\mathrm j}>$1.5~GeV)
and the magnitude of the sum of the track charges 
($|Q_{\mathrm j}|$) must not exceed 1.
The $P_{\mathrm t}/\Ebeam$ distributions for region (III) are
shown in Figure~1(c) after these cuts.
In region (I), if the acoplanarity angle is smaller than 70\degree,
cuts are applied on 
$P_{\mathrm t}$, $a_{\mathrm t}$ (the 
transverse momentum perpendicular to the event thrust axis),
and $|\cos \theta_a|$, 
where $\theta_a \equiv \tan^{-1}(a_{\mathrm t}/P_{\mathrm{z}})$.
These cuts reduce the background contamination from 
two-photon and $\tautau$ processes.
They are identical to those in Ref.~\cite{LEP18-opal} for region
(Ib) but have been further optimized for region (Ia) where
$a_{\mathrm t}/E_{\mathrm{beam}}$ is required to lie between 0.025 and
0.1 and $|\cos \theta_a|$ to be smaller than 0.9.
To further reduce the two-photon background, cuts on 
$P_{\mathrm t}$ and $| \cos\theta_{\mathrm{miss}}|$
are applied in region (I) if the acoplanarity angle is larger than 70\degree\@ 
and in all other regions for any value of acoplanarity angle.
The cuts on $| \cos\theta_{\mathrm{miss}}|$ are identical to those 
in Ref.~\cite{LEP18-opal} apart from region (Ia) where it is required to be 
smaller than 0.75. 
$P_{\mathrm t}/E_{\mathrm{beam}}$ should be between 0.02 and 0.04, 
between 0.03 and 0.05, and between 0.035 and 0.075 for regions (Ia), (Ib) 
and (II), and should be larger than 0.095 and 0.1 for regions (III) and (IV).

To reduce the background from $\ee\mumu$ events in which one of the muons 
is emitted at a small polar angle and is not reconstructed as a good
track, events are rejected if there is a track segment in the
muon chamber or a hadron calorimeter cluster at a small polar angle,
and within 1~radian in ($r$,$\phi$) of the 
missing momentum direction ($\vec P_{\mathrm{miss}}$).
%  in the plane 
% perpendicular to the beam axis. 

Cuts on $|\cos \theta_{\mathrm j}|$ and $\phi_{\mathrm{acop}}$
are applied to reject two-photon, lepton-pair and 
$\gamma^* \Z^{(*)} \ra \ell^+ \ell^- \nu \bar{\nu}$
events. The values of
these cuts have been slightly modified with respect to
Ref.~\cite{LEP18-opal} in that 
$|\cos \theta_{\mathrm j}|$ is now required to be smaller than 0.75 in
region (Ia), and $\phi_{\mathrm{acop}}$ should lie between 50\degree\@
and 160\degree\@ for region (Ia) and between 20\degree\@ and 160\degree\@
for region (Ib).
$\WW \ra \ell^+ \nu \ell^- \bar{\nu}$ events are rejected by
upper cuts on $M_{\mathrm{vis}}$ (20~GeV, 25~GeV, 30~GeV, 50~GeV and
75~GeV for regions (Ia) to (IV)) and
on the higher energy of the two jets, $E_1 / E_{\mathrm{beam}}$
(identical to Ref.~\cite{LEP18-opal} apart from region (Ia) where it
is required to be smaller than 0.2)\@.

The numbers of background events expected from
the four different sources, for 
each $\Delta M_+$ region, are given in Table~\ref{tab:cutCch}.
The typical detection efficiencies for $\chp_1 \chm_1$ events 
are 30--65\% for $\Delta M_+ \gsim 10$~GeV,
and the modest efficiency of 20\% is obtained for $\Delta M_+=5$~GeV\@.

\begin{table}[ht]%-----------------------------------------------
\centering
\begin{tabular}{|l||c|c|c|c|c|}
\hline%-----------------------------------------------------------------
Region  & \multicolumn{2}{|c|}{I} & II & III &  IV  \\ 
\cline{1-3}  
Sub-Region                    &
\multicolumn{1}{|c|}{a} & \multicolumn{1}{|c|}{b} & 
$_{10~{\mathrm{GeV}} < \Delta M_+} $ &     
$_{m_{\chp_1}/2 < \Delta M_+} $ &
$_{m_{\chp_1}-20~{\mathrm{GeV}} < \Delta M_+} $ \\
               &  
$_{3  \leq \Delta M_+ \leq 5~{\mathrm{GeV}}}$  &
$_{5  < \Delta M_+ \leq 10~{\mathrm{GeV}}}$ &
$_{\leq m_{\chp_1}/2}$  &
$_{\leq m_{\chp_1}-20~{\mathrm{GeV}}}$  &
$_{\leq m_{\chp_1}}$   \\
\hline%-----------------------------------------
\hline%-----------------------------------------------------------------
background & \multicolumn{5}{|c|}{ }   \\  
\hline% ----------------------------------------
$\gamma \gamma$     & 6.89 & 3.31 & 9.07 & 13.91 & 12.99  \\  
\hline% ----------------------------------------
$\ellell (\gamma)$  & 0.02 & 0.16 & 1.10 & 2.97 & 5.39  \\  
\hline% ----------------------------------------
$\qq (\gamma)$      & 0.00 & 0.00 & 0.04 & 0.04 & 0.04 \\  
\hline% ----------------------------------------
4f                  & 0.95 & 1.89 & 9.72 & 54.96 & 97.97 \\  
\hline% ----------------------------------------
\hline% ----------------------------------------
total bkg     &  7.9$\pm$1.1& 5.4$\pm$0.8& 20.0$\pm$1.4
              & 71.9$\pm$2.1 & 116.4$\pm$2.2 \\     
\hline% ----------------------------------------
\hline% ----------------------------------------
observed       & 12  & 8  &  23  &  73  &  112  \\  
\hline% ----------------------------------------
\end{tabular}          
\caption[]{%\sl
  \protect{\parbox[t]{15cm}{
The numbers of expected background events 
for various Standard Model processes, 
normalised to the integrated luminosity, 
and the number of observed 
events in each $\Delta M_+$ region for category (C).
The errors in the total background include only the statistical error
of the Monte Carlo samples.
}} }
\label{tab:cutCch}
\end{table}%------------------------------------------------------------

\subsection{Detection of neutralinos}

To obtain optimal performance, the event sample is divided 
into two mutually exclusive categories, motivated
by the topologies expected for neutralino events. 
\begin{itemize}
\item[(C)] $N_{\mathrm{ch}} \leq 4$.
Signal events in which $\nt_2$ decays into $\nt_1 \ellell$ tend
to fall into this category.
Also, when the mass difference 
between $\nt_2$ and $\nt_1$ ($\Delta M_0 \equiv m_{\nt_2} -
m_{\nt_1}$) is small,
signal events tend to fall into this category independently of the
$\nt_2$ decay channel.
\item[(D)] $N_{\mathrm{ch}} > 4$. Signal events
in which $\nt_2$ decays into $\nt_1 {\mathrm q}\bar{\mathrm q}$
tend to fall into this category for modest and large values of 
$\Delta M_0$\@.
\end{itemize}

The event topology of $\nt_1 \nt_2$ events depends mainly on the 
difference between the $\nt_2$ and $\nt_1$
masses. Separate selection criteria are therefore used in four
$\Delta M_0$ regions:  
(i) $\Delta M_0 \leq 10$~GeV, 
(ii) $10 < \Delta M_0 \leq 30$~GeV,
(iii) $30 < \Delta M_0 \leq 80$~GeV,
(iv) $\Delta M_0 > 80$~GeV\@.
In regions (i) and (ii), the main sources of background are 
two-photon and $\gamma^* \Z^{(*)} \ra \qq \nunu$ processes.
In regions (iii) and (iv), the main sources of background are
four-fermion processes ($\WW$, $\Wenu$ and $\gamma^* \Z^{(*)}$).
The fraction of events falling into
category (C) is 10-20\% for $\Delta M_0 \geq 20$~GeV but
increases to about 70\% when  $\Delta M_0 \leq 5$~GeV.
The fraction of invisible events due to $\nt_2 \ra \nt_1\Zrv \ra \nt_1 \nunu$ 
decays is 20-30\% depending on $\Delta M_0$\@.

The selection criteria applied for the low-multiplicity events (category (C))  
in regions (i), (ii), (iii)  and (iv) are identical to
those used in analysis (C) of the chargino search
in regions (Ia), (II), (III) and (IV), 
respectively (see Table~\ref{tab:cutCch}). 

Events falling into category (D) have typically a monojet or a di-jet 
topology with large missing transverse momentum.
The cuts described below are applied for these topologies. They are
identical to the ones of Ref.~\cite{LEP18-opal} unless otherwise stated.

%=======================================================================
\subsubsection{Analysis (D) (\boldmath$N_{\mathrm{ch}} > 4$)}
%=======================================================================

To reduce the background from two-photon and $\qq(\gamma)$ processes, 
cuts on $|\cos(\thmiss)|$, $E_{\mathrm{fwd}}/E_{\mathrm{vis}}$ 
and missing transverse momenta are applied.
The acoplanarity angle
should be large to remove $\qq(\gamma)$ background events.
To ensure the reliability of the measurement of $\phi_{\mathrm{acop}}$,
both jets should have a polar angle $\theta_{\mathrm j}$ satisfying 
$|\cos\theta_{\mathrm j} |<0.95$.
In region (iv), the $\phi_{\mathrm{acop}}$ cut is loosened
with respect to regions (i)--(iii), 
since the acoplanarity angle of signal events becomes smaller. 
To compensate for the resulting higher $\qq(\gamma)$ background, 
the $E_{\mathrm{fwd}}/E_{\mathrm{vis}}$ cut 
is tightened.

After these cuts, the remaining background events come 
predominantly from
$\gamma^* \Z^{(*)} \ra \qq \nunu$, $\WW \ra \ell \nu\qq^{'}$
and $\Wenu \ra  \qq ^{'}{\mathrm e} \nu$.
Cuts on the visible mass are applied to reduce $\WW \ra \ell \nu\qq^{'}$
and $\Wenu \ra  \qq ^{'}{\mathrm e} \nu$ processes.
$\gamma^* \Z^{(*)} \ra \qq \nunu$  background events are removed by a  
cut on the ratio of the visible mass to the visible energy, which is required
to be larger than 0.3 for
region (i) and 0.25 for regions (ii) and (iii).
In regions (iii) and (iv), 
$d_{23}^2 \equiv y_{23} E_{\mathrm{vis}}^2 < 30$~GeV$^2$ is required  
to select a clear two-jet topology and to reject 
$\WW \ra \tau \nu \qq'$ events.
In Figure~1(d) the $d_{23}^2$ 
distribution is shown for region (iv)
after all the other cuts. 
The numbers of background events expected from
the four different sources, for 
each $\Delta M_0$ region, are given in Table~\ref{tab:cutDnt}.
Typical detection efficiencies for $\nt_2 \nt_1$ events are 50--65\%
for $\Delta M_0 > 10$~GeV\@.

\begin{table}[htb]%------------------------------------------------------
\centering
\begin{tabular}{|l||c|c|c|c|}
\hline%-----------------------------------------------------------------
Region  & i & ii & iii & iv \\
 &
$_{\Delta M_0 \leq 10~{\mathrm{GeV}}}$  &
$_{10~{\mathrm{GeV}} < \Delta M_0 \leq 30~{\mathrm{GeV}}}$  &
$_{30~{\mathrm{GeV}} < \Delta M_0 \leq 80~{\mathrm{GeV}}}$  &
$_{\Delta M_0 > 80~{\mathrm{GeV}}}$  \\
\hline%-----------------------------------------------------------------
\hline%-----------------------------------------
background & \multicolumn{4}{|c|}{ }      \\  
\hline% ----------------------------------------
$\gamma \gamma$     & 3.06 & 4.91 & 1.48 & 0.00 \\ 
\hline% ----------------------------------------
$\ellell (\gamma)$  & 0.00 & 0.07 & 0.13 & 0.33  \\  
\hline% ----------------------------------------
$\qq (\gamma)$      & 0.00 & 0.00 & 0.00 & 0.04  \\  
\hline% ----------------------------------------
4f                  & 0.25 & 4.62 & 11.72 & 39.97 \\  
\hline% ----------------------------------------
\hline% ----------------------------------------
total bkg & 3.3$\pm$1.1 & 9.6$\pm$1.5 & 13.3$\pm$1.0 
              & 40.3$\pm$1.2 \\  
\hline% ----------------------------------------
\hline% ----------------------------------------
observed       & 2  &  8  &  14 & 39   \\  
\hline% ----------------------------------------
\end{tabular}
\caption[]{%\sl
  \protect{\parbox[t]{15cm}{
The numbers of expected background events 
for various Standard Model processes, 
normalised to the integrated luminosity, 
and the number of observed 
events in each $\Delta M_0$ region for category (D).
The errors in the total background include only the statistical error
of the Monte Carlo samples.
}} 
}
\label{tab:cutDnt}
\end{table}%------------------------------------------------------------

%=======================================================================
\section{Systematic uncertainties}
%=======================================================================
Systematic uncertainties on the number of expected signal  and
background events are  estimated in the same manner as
in the previous papers~\cite{LEP17-opal} and~\cite{LEP18-opal} and
are only briefly described here.

For the number of expected signal events the uncertainties arise from
the measurement of the integrated luminosity (0.5\%), Monte-Carlo
statistics and interpolation of the efficiencies to arbitrary values
of $m_{\chpm_1}$ and $m_{\nt_1}$ (2--10\%), modelling of the cut
variables in the Monte Carlo simulations (4--10\%), fragmentation 
uncertainties in hadronic decays ($<2$\%)
and detector calibration effects ($<1$\%).

The angular distributions of the chargino and neutralino final-state decay
products and
their effect on the resulting signal detection efficiencies  
depend on the details of the parameters of the Constrained MSSM~\cite{mssm}.
However, the corresponding 
variation of the efficiencies is determined to be less than
5\% (relative),
and this is taken into account
in the systematic errors when obtaining the limits.
Consequently, the limits are independent
of the details of the Constrained MSSM.

For the expected number of background events, the uncertainties are due
to Monte Carlo statistics (see
Tables~1 to~4), uncertainties in the amount of two-photon background
(30\%), uncertainties in the simulation of the four-fermion processes
(17\%), and modelling of the cut variables ($<7$\%), as determined in 
Ref.~\cite{LEP17-opal}.

In addition to effects included in the detector simulation, an
efficiency loss of 2.9\% (relative) arises from beam-related
background in the silicon-tungsten forward calorimeter and 
in the forward detector which
is estimated using random beam crossing events.

%=======================================================================
\section{Results}
%=======================================================================

No evidence for $\chp_1 \chm_1$ and $\nt_1 \nt_2$ production 
is observed.
Exclusion regions and limits are determined by
using the likelihood ratio method~\cite{LH}\@,
which assigns 
greater weight to the analysis which has the largest sensitivity.  
Systematic uncertainties on the efficiencies and on the number of expected
background events are taken into account in the cross-section 
limit calculations 
according to Ref.~\cite{Cousins}. 

\subsection{Limits on the $\chp_1 \chm_1$ 
and $\nt_2 \nt_1$ production cross-sections}

Figures~2(a) and (b) show model-independent upper-limits (95\% C.L.) 
on the production cross-sections of $\chp_1 \chm_1$ and 
$\nt_2 \nt_1$, respectively.
These are obtained assuming
the specific decay mode $\chpm_1 \ra \nt_1 {\mathrm W}^{(*)\pm}$ 
for $\chp_1 \chm_1$,
and $\nt_2 \ra \nt_1 {\mathrm Z}^{(*)}$ 
for $\nt_1 \nt_2$ production.
The results from the $\roots =$183~GeV~\cite{LEP18-opal} analysis 
are also included in the limit calculation~\footnote{
In calculating limits, cross-sections
at different $\roots$ were estimated
by weighting by $\bar\beta / s$,
where $\bar\beta$ is 
$p_{\chpm_1}/E_{\mathrm{beam}}$ for $\chp_1 \chm_1$ production 
or  $p_{\nt_2}/E_{\mathrm{beam}} = p_{\nt_1}/E_{\mathrm{beam}}$ for 
$\nt_2 \nt_1$ production.}\@.

If the cross-section for $\chp_1 \chm_1$ is larger than
0.75~pb and $\Delta M_+$ is between 5~GeV and about 80~GeV,
$m_{\chp_1}$ is excluded at the 95\% C.L. up to the kinematic limit.
%, assuming $Br(\chp_1 \ra \nt_1 {\mathrm W}^{(*)+}) = 100\%$\@.
If the cross-section for $\nt_2 \nt_1$ is larger than
0.95~pb and $\Delta M_0$ is greater than 7~GeV, 
$m_{\nt_2}$ is excluded up to
the kinematic limit at 95\% C.L.
% assuming $Br(\nt_2 \ra \nt_1 {\mathrm Z}^{(*)}) = 100\%$\@.

\subsection{Limits in the MSSM parameter space}

% The results of the above searches can be interpreted within the
% framework of the Constrained MSSM (CMSSM)~\cite{mssm}\@.
The phenomenology of the gaugino-higgsino
sector of the MSSM is mostly determined by the following 
parameters: the SU(2) gaugino mass parameter at the weak scale
($M_2$), 
the mixing parameter of the two Higgs doublet fields ($\mu$) and the
ratio of the vacuum expectation values of the two Higgs doublets ($\tan\beta$).
% In the absence of light sfermions~\cite{LEP18-slepton} and a
% light SUSY Higgs~\cite{LEP18-higgs},
Assuming sfermions~\cite{LEP18-slepton} and SUSY Higgs~\cite{LEP18-higgs} 
sufficiently heavy not to intervene in the 
decay channels,
these three parameters are sufficient to describe
the chargino and neutralino sectors completely.  
Within the Constrained MSSM~\cite{mssm}, 
a large value of the common 
scalar mass, $m_0$ (e.g., $m_0 = 500$~GeV) leads to heavy sfermions
and therefore to a negligible 
suppression of the cross-section due to interference 
with $t$-channel sneutrino exchange. Chargino 
decays would then proceed predominantly
via a virtual or real W.
On the other hand, a light $m_0$ results in a low value of
the mass of the sneutrino,
enhancing the contribution of the $t$-channel exchange diagrams 
that have destructive interference with $s$-channel diagrams,
thus reducing the cross-section for chargino pair production.
Small values of $m_0$ also enhance the leptonic branching
ratio of charginos.
%Certain values of $m_0$ can lead to the condition $m_{\snu} < m_{\chpm_1}$
%and result in the two-body decay mode $\chpm_1 \ra \snu \ell^{\pm}$
%being dominant.  The chargino detection efficiency can be small
%or zero for these decays, 
%particularly when $m_{\snu} \approx m_{\chpm_1}$, 
%leading to severe degradation in sensitivity.

From the input parameters $M_2$, $\mu$, $\tan \beta$,
$m_0$ and $A$ (the trilinear Higgs coupling),
masses, production cross-sections and
branching fractions are calculated
according to the Constrained MSSM~\cite{mssm,chargino,neutralino,mssm2}.
For each set of input parameters, the total numbers of
$\chp_1 \chm_1$, $\nt_2 \nt_1$, $\nt_3 \nt_1$ and $\nt_2 \nt_2$ events
expected to be observed are calculated using
the integrated luminosity, the cross-sections,
the branching fractions, and 
the detection efficiencies (which depend upon the masses of the chargino,
the lightest neutralino and next-to-lightest neutralino).
Contributions from channels such as $\nt_1 \nt_4$, $\nt_2
\nt_4$, etc\ldots are not included.
The $\nt_3 \nt_1$ channel is similar to the $\nt_2 \nt_1$ channel,
and cascade decays through $\nt_2$ are taken into account.
The relative importance of each of the analyses (A)--(D) changes with
the leptonic or hadronic branching ratios, and the likelihood
ratio method~\cite{LH} is used to optimally weight each analysis
depending on these branching ratios.

Results are presented for two cases: (i) $m_0 = 500$~GeV (i.e., heavy 
sfermions), 
and (ii) the value of $m_0$
that gives the smallest total numbers of expected chargino
and neutralino events 
taking into account cross-sections,
branching ratios, and detection efficiencies
for each set of
values of $M_2$, $\mu$ and $\tan \beta$.
This latter value of $m_0$ leads to the most conservative limit at that point,
so that the resulting limits are valid for all $m_0$.
In searching for this value of $m_0$, 
only those values are considered that 
are compatible with the current limits
on the sneutrino mass ($m_{\snu_L} > 43$~GeV~\cite{PDG98}), and
upper limits on the cross-section for slepton pair
production, particularly
right-handed smuon and selectron pair 
production~\cite{LEP18-slepton}\@.
Particular attention is paid to the region of values of $m_0$ 
leading to the mass condition
$m_{\snu} \approx m_{\chpm_1}$ by taking finer steps in the value of $m_0$\@.
Note that we assume the stau mixing angle to be zero, and
the added complication of possible enhanced decays to third-generation
particles at large values of $\tan\beta$ because of stau mixing is
ignored.
When $m_{\snu} \leq m_{\chpm_1}$, resulting in a topology of 
acoplanar leptons and missing momentum, the upper limits
on the cross-section for the two-body chargino decay from
Ref.~\cite{LEP18-slepton} are used, 
while for $m_{\snu} > m_{\chpm_1}$ the three-body decays are dominant.
%The contribution of the cascade decays $\chp_1 \ra \nt_2 \mathrm{X}$ followed by
%$\nt_2 \ra \nt_1 \mathrm{Y}$ were also included.
Single photon topologies from $\nt_2 \nt_1$ production and acoplanar photons
with missing energy topologies from $\nt_2 \nt_2$ with 
photonic decay $\nt_2 \ra \nt_1 \gamma$ 
are taken into account
using the 95\% C.L. cross-section upper limits on these topologies from
OPAL results~\cite{gamgam}.
In both of these cases, if the relevant product of
cross-section and branching ratio for a particular set
of MSSM parameters is greater than the measured 95\% C.L. upper limit
presented in that paper, then that set of parameters is considered to be
excluded.

The following regions of the Constrained MSSM parameters are scanned:
$0\le M_2 \le 2000$~GeV,
$|\mu| \le 500$~GeV, and
$A = \pm M_2, \,\, \pm m_0$ and 0.
The typical scan step is 0.2~GeV.
Extensions beyond the scanned range have negligible effect on the quoted 
limits.
The choice of $A$ values is related to various scenarios of stop mixing, 
influencing the Higgs sector but having essentially no effect on the gaugino 
sector. No significant dependence on $A$ is observed.
Figure~3 shows the resulting exclusion regions 
in the ($M_2$,$\mu$) plane
for $\tan\beta = 1.5$ and 35 with $m_0 \geq 500$~GeV and for all $m_0$.

The restrictions on the Constrained MSSM parameter space presented here can be
transformed into exclusion regions in the ($m_{\chpm_1}$,$m_{\nt_1}$)
or ($m_{\nt_2}$,$m_{\nt_1}$) plane.
A given mass pair is excluded only if
all considered Constrained MSSM parameters in the scan which lead
to that same mass pair are
excluded at the 95\%~C.L.
The $\chpm_1$ mass limits are summarised 
in Table~\ref{tab:resultsc}. 
In the ($m_{\chpm_1}$,$m_{\nt_1}$) plane, Figures~4(a) and (b)
show the corresponding
95\% C.L. exclusion regions for $\tan\beta = 1.5$ and 35.
Figures~4(c) and (d) show the corresponding
95\% C.L. exclusion regions in the ($m_{\nt_2}$,$m_{\nt_1}$) 
plane, for $\tan\beta = 1.5$ and 35.
Mass limits on $\nt_1$, $\nt_2$, and $\nt_3$ 
are summarised in Table~\ref{tab:resultsn}.

\begin{table}[htb]%------------------------------------------------------
\centering
\begin{tabular}{|c|c||c|c|}
\hline
 &     & $\tan \beta = 1.5$  & $\tan \beta = 35$ \\
\hline%-----------------------------------------------------------------
\hline%-----------------------------------------------------------------
$m_0 \ge 500$~GeV &
$\Delta M_+ \geq 5$~GeV     & $m_{\chp_1}>93.6$~GeV & $m_{\chp_1}>94.1$~GeV \\
\hline%-----------------------------------------------------------------
All $m_0$ (see text) &
$\Delta M_+ \geq 5$~GeV    & $m_{\chp_1}>78.0$~GeV & $m_{\chp_1}>71.7$~GeV \\
\hline%----------------------------------------------------------------------
\end{tabular}
\caption[]{%\sl
   \protect{\parbox[t]{15cm}{
Lower limits at 95\% C.L. obtained on the lightest chargino mass.
}} }
\label{tab:resultsc}
\end{table}%------------------------------------------------------------

\begin{table}[htb]%------------------------------------------------------
\centering
\begin{tabular}{|c|c||c|c|}
\hline
   &   & $\tan \beta = 1.5$  & $\tan \beta = 35$ \\
\hline%-----------------------------------------------------------------
\hline%-----------------------------------------------------------------
$m_0 \ge 500$~GeV & No $\Delta M_0$ restriction& 
       $m_{\nt_1}>40.2$~GeV & $m_{\nt_1}>48.5$~GeV \\ \cline{2-4}
 & $\Delta M_0 \geq 10$~GeV   & $m_{\nt_2}>67.8$~GeV & $m_{\nt_2}>94.3$~GeV \\
 &                            & $m_{\nt_3}>106.0$~GeV  & $m_{\nt_3}>124.0$~GeV \\
\hline%-----------------------------------------------------------------
All $m_0$ (see text) & No $\Delta M_0$ restriction& 
       $m_{\nt_1}>34.1$~GeV & $m_{\nt_1}>38.9$~GeV \\ \cline{2-4}
 & $\Delta M_0 \geq 10$~GeV   & $m_{\nt_2}>55.9$~GeV & $m_{\nt_2}>86.2$~GeV \\
 &                            & $m_{\nt_3}>106.6$~GeV  & $m_{\nt_3}>134.3$~GeV \\
\hline%-----------------------------------------------------------------
\end{tabular}
\caption[]{%\sl
   \protect{\parbox[t]{15cm}{
Lower limits at 95\% C.L. obtained on $m_{\nt_1}$, $m_{\nt_2}$, and
$m_{\nt_3}$. 
}} }
\label{tab:resultsn}
\end{table}%------------------------------------------------------------

Figure~5 shows the dependence of the mass limits on 
the value of $\tan\beta$.  Of particular interest is the absolute
lower limit, in the framework of the Constrained MSSM, 
on the mass of the lightest 
neutralino of $m_{\nt_1} > 32.8$~GeV (31.6~GeV) at 95\% C.L. 
for $m_0 \ge 500$~GeV (all $m_0$).
This has implications for direct searches 
for the lightest neutralino as a candidate for dark matter~\cite{dark}.
Since the formulae for couplings and masses in the gaugino sector
are symmetric in $\tan\beta$ and $1/\tan\beta$, these results
also hold for $\tan\beta < 1$.

%=======================================================================
\section{Summary and Conclusion}
%=======================================================================

A data sample corresponding to
an integrated luminosity of 182.1~pb$^{-1}$
at $\roots = $188.6~GeV,
collected with the OPAL detector, has been analysed
to search for pair-production of charginos ($\chp_1\chm_1$) and neutralinos  
($\nt_2\nt_1$) predicted by supersymmetric theories. 
Decays of $\chpm_1$ into 
$\nt_1 \ell^\pm \nu$ or $\nt_1 \mathrm{q} \mathrm{\overline{q}'}$
and decays of $\nt_2$ into $\nt_1 \nunu$, $\nt_1 \ellell$ or
$\nt_1 {\mathrm q}\bar{\mathrm q}$ are looked for.
No evidence for such events has been observed.
The exclusion limits on $\chpm_1$ and $\nt_j$ production are significantly 
higher with respect to the results obtained at lower centre-of-mass energies.

Exclusion regions valid at the 95\% confidence level
have been derived in the framework of the Constrained MSSM, in which 
only three parameters ($M_2$, $\mu$ and $\tan\beta$) are necessary to 
describe the chargino and neutralino sectors.
These restrictions in parameter space have been transformed into 
mass limits valid at the 95\% confidence level.
Assuming $m_{\chpm_1}-m_{\nt_1} \geq 5$~GeV,
the lower mass limit of the chargino is 93.6~GeV
for $\tan \beta = 1.5$ and 94.1~GeV for $\tan \beta = 35$
for the case of a large universal scalar mass ($m_0 \geq$~500~GeV);
for all $m_0$, the mass limit 
is 78.0~GeV for $\tan \beta = 1.5$
and 71.7~GeV for $\tan \beta = 35$\@.
The lower mass limit for the lightest neutralino is 32.8~GeV for the case of
$m_0 \ge 500$~GeV and
31.6~GeV for all $m_0$.
This latter result has implications for searches for 
the lightest neutralino as a dark matter candidate.

\bigskip
%-----------------------------------------------------------------------
\section*{Acknowledgements}
%-----------------------------------------------------------------------
We particularly wish to thank the SL Division for the efficient operation
of the LEP accelerator at all energies
 and for their continuing close cooperation with
our experimental group.  We thank our colleagues from CEA, DAPNIA/SPP,
CE-Saclay for their efforts over the years on the time-of-flight and trigger
systems which we continue to use.  In addition to the support staff at our own
institutions we are pleased to acknowledge the  \\
Department of Energy, USA, \\
National Science Foundation, USA, \\
Particle Physics and Astronomy Research Council, UK, \\
Natural Sciences and Engineering Research Council, Canada, \\
Israel Science Foundation, administered by the Israel
Academy of Science and Humanities, \\
Minerva Gesellschaft, \\
Benoziyo Center for High Energy Physics,\\
Japanese Ministry of Education, Science and Culture (the
Monbusho) and a grant under the Monbusho International
Science Research Program,\\
Japanese Society for the Promotion of Science (JSPS),\\
German Israeli Bi-national Science Foundation (GIF), \\
Bundesministerium f\"ur Bildung, Wissenschaft,
Forschung und Technologie, Germany, \\
National Research Council of Canada, \\
Research Corporation, USA,\\
Hungarian Foundation for Scientific Research, OTKA T-029328, 
T023793 and OTKA F-023259.\\
%=======================================================================
%       References
%=======================================================================

%\newpage

%=======================================================================
%       Figures
%=======================================================================
\newpage
\begin{figure}
\begin{minipage}{16.cm}
\begin{center}\mbox{
\epsfig{file=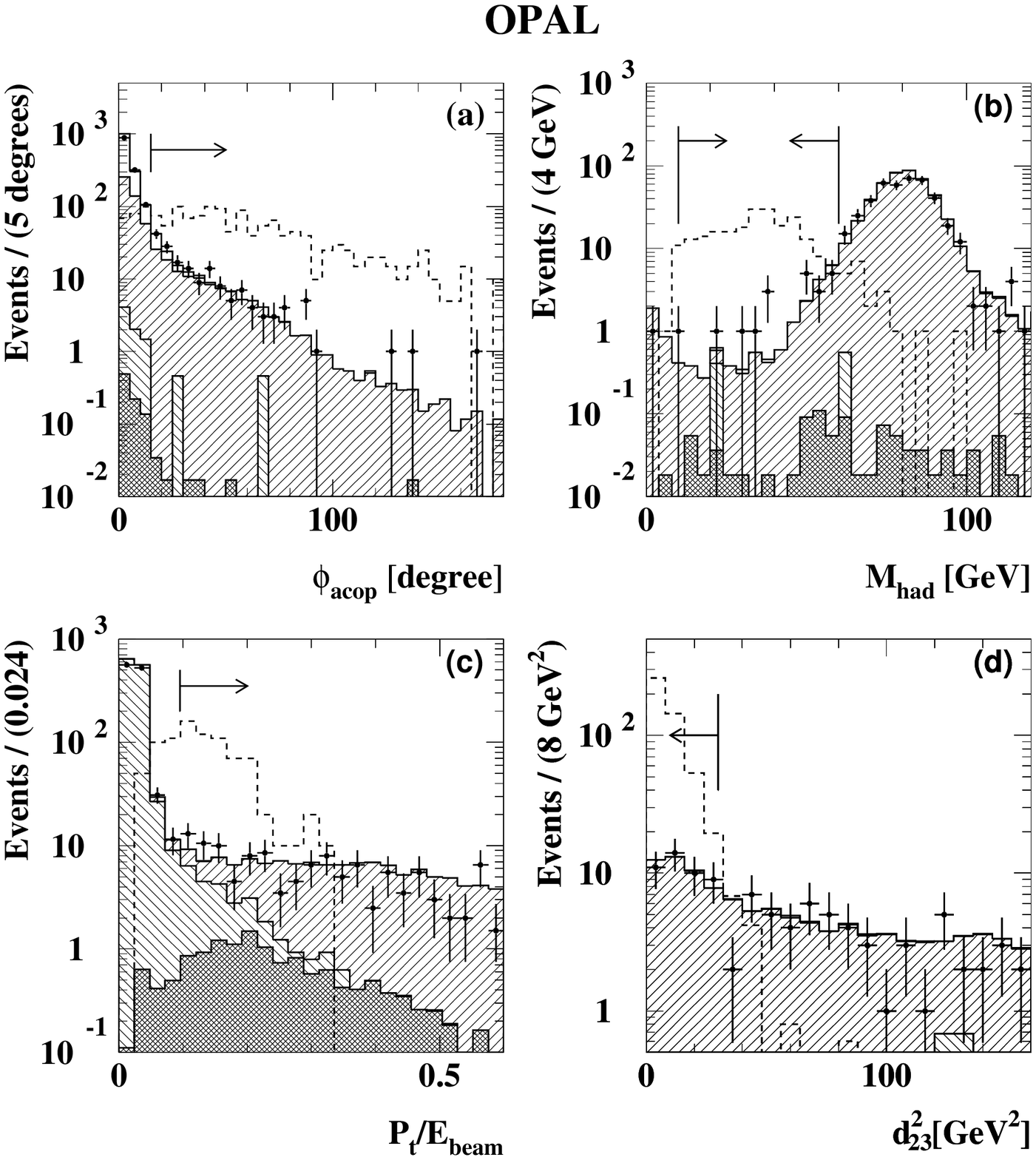, width=16cm}
% \epsfig{file=fig1b.eps, height=8cm} 
% }\end{center}
% \vspace*{-1.mm}
% \begin{center}\mbox{
% \epsfig{file=fig1c.eps, height=8cm}
% \epsfig{file=fig1d.eps, height=8cm}
}\end{center}
\vspace*{-1.mm}
\caption[1]
{
Distributions of some essential observables used to select chargino
and neutralino events: 
(a) acoplanarity angle in analysis (A),
(b) invariant mass of the event calculated excluding
the highest momentum isolated lepton in analysis (B),
(c) ratio of transverse momentum to beam energy, 
$P_{\mathrm t}/\Ebeam$, in analysis (C), and
(d) $d_{23}^2$ in analysis (D)\@.
The data are shown with 
error bars and the distributions from
the background processes are shown as filled histograms: dilepton events
(double hatched area), two-photon processes 
(negative slope hatching area), four-fermion processes
including W-pair events (positive slope hatching area), and
multihadronic events (open area).
The arrows, pointing into the accepted regions, 
show where the analysis cuts are applied.
In Figures (a)-(c) the dashed line shows the prediction
for a chargino signal
with $m_{\chp_1} = 94$~GeV and $m_{\nt_1} = 47$~GeV.
In Figure (d) the dashed line shows the prediction for a neutralino signal
with $m_{\nt_2} = 145$~GeV and $m_{\nt_1} = 35$~GeV.
The normalisations of the signal histograms are arbitrary.
}
\label{fig1}
\end{minipage}
\end{figure}
%-----------------------------------------------------------------------
\newpage
\begin{figure}
\begin{minipage}{16.cm}
\vspace{-15.mm}
\begin{center}\mbox{
\epsfig{file=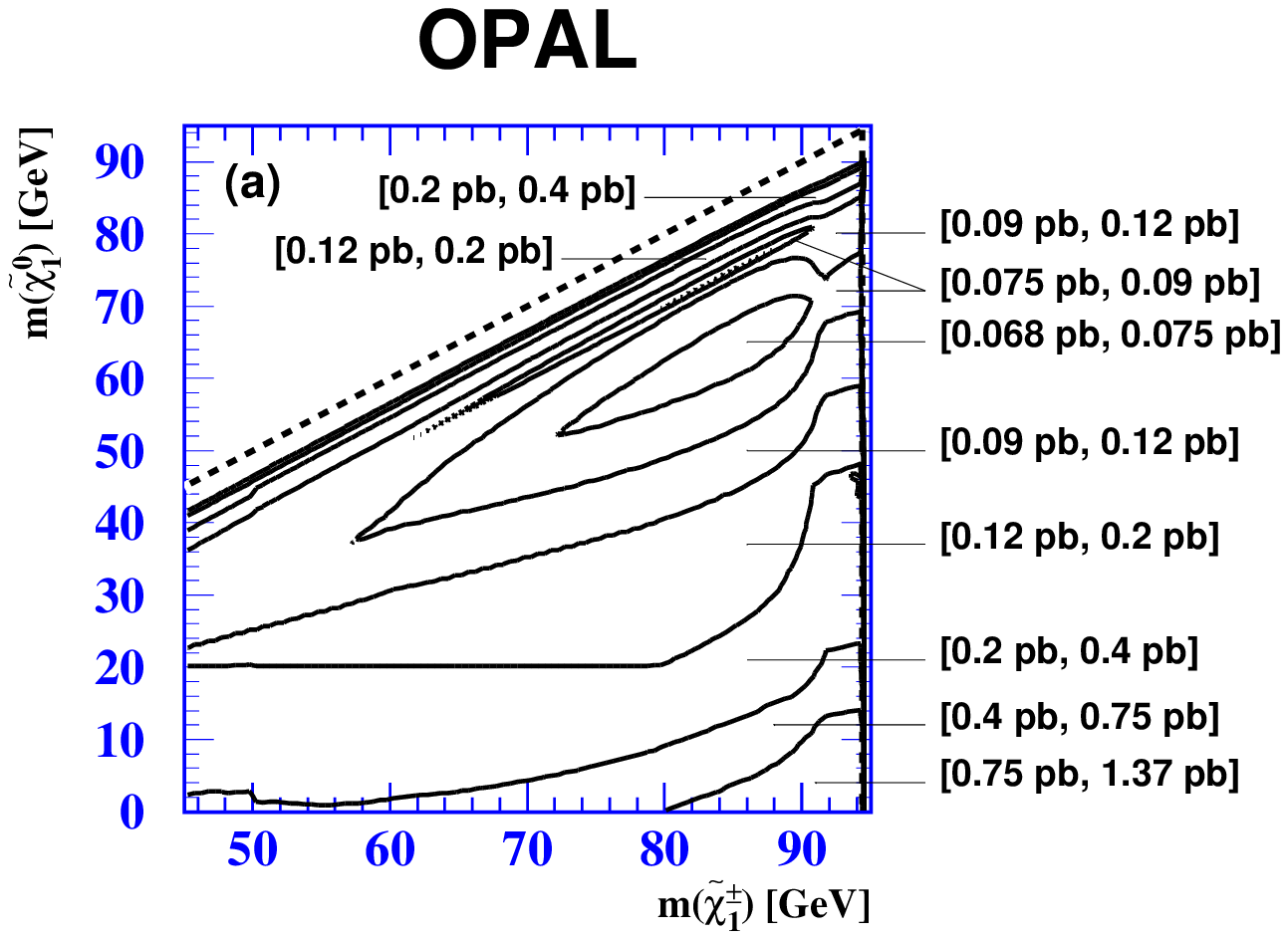, width=14.0cm}
}\end{center}
\begin{center}\mbox{
\epsfig{file=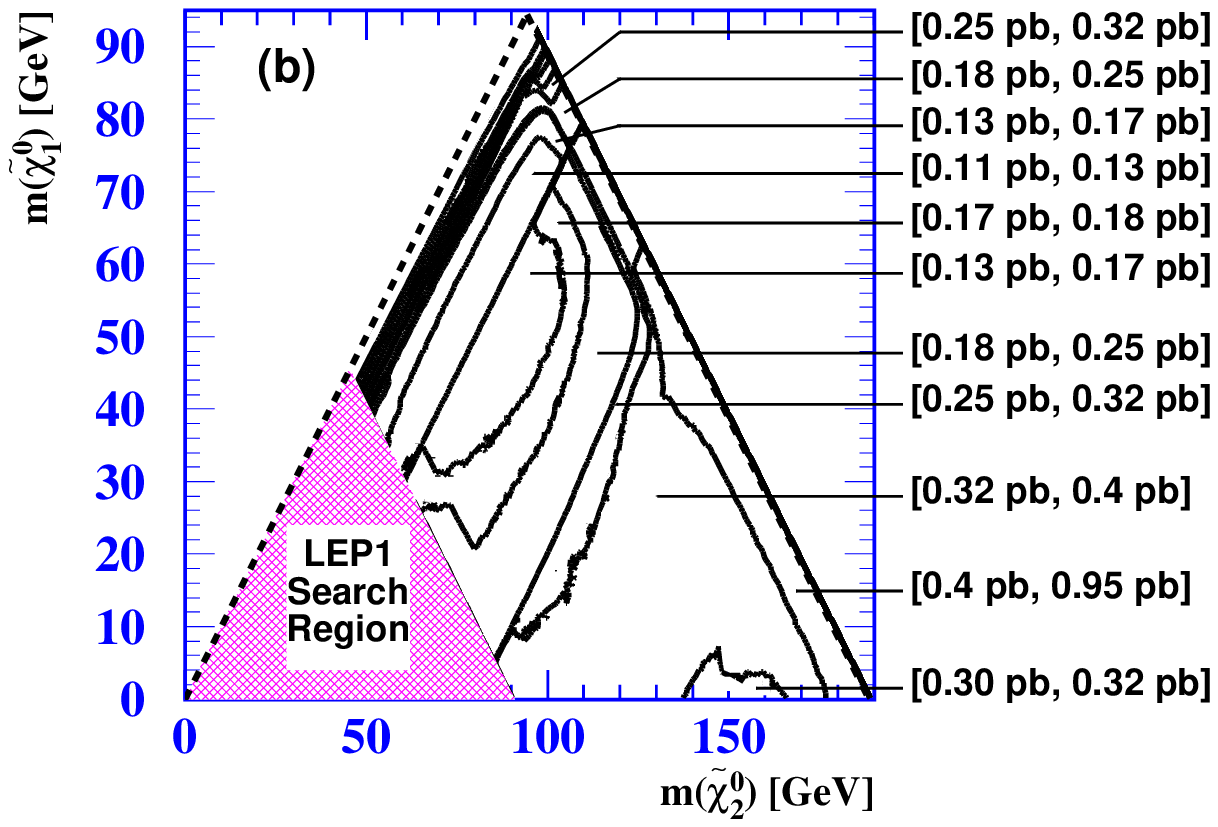, width=14.0cm}
}\end{center}
\vspace{-9mm}
\caption[1]
{
Contour plots of the 95\% C.L. upper-limits on the production
cross-sections at $\roots$ = 189~GeV 
(a) $\sigma_{\chp_1 \chm_1}$ for $\ee \ra \ \chp_1 \chm_1$  
and (b) $\sigma_{\nt_2 \nt_1}$ for $\ee \ra \ \nt_1 \nt_2$.
The $\chp_1$ is assumed to decay only into $\nt_1 {\mathrm W}^{(*)\pm}$, and
the $\nt_2$ is assumed to decay only into $\nt_1 {\mathrm Z}^{(*)}$.
The cross-hatched region, for which 
$m_{\nt_1}+m_{\nt_2}<m_{{\mathrm Z}}$, 
is covered by searches at LEP1, which set upper-limits on 
$\sigma_{\nt_2 \nt_1}$ of order 5~pb, and is not considered in this
analysis. 
The kinematical boundaries for $\chp_1 \chm_1$ and $\nt_1 \nt_2$ 
are shown as dashed lines.
If $\sigma_{\chp_1 \chm_1}$ is smaller than
0.068~pb, there is no exclusion region in the ($m_{\chp_1}$,$m_{\nt_1}$)
plane. If $\sigma_{\chp_1 \chm_1}$ is between 0.068 and 0.075~pb a
small region is excluded around $m_{\chp_1}=85$~GeV and
$m_{\nt_1}=65$~GeV. 
Similarly, if $\sigma_{\nt_2 \nt_1}$ is 
smaller than 0.11~pb no region can be excluded in the
($m_{\nt_2}$,$m_{\nt_1}$) plane. If $\sigma_{\nt_2 \nt_1}$ is between
0.11 and 0.13~pb a small region is excluded around
$\Delta M_0 \equiv m_{\nt_2} -
m_{\nt_1} = 25$~GeV.

}
\end{minipage}
\end{figure}
%-----------------------------------------------------------------------
\newpage
\begin{figure}
\begin{minipage}{16.cm}
\begin{center}\mbox{
\epsfig{file=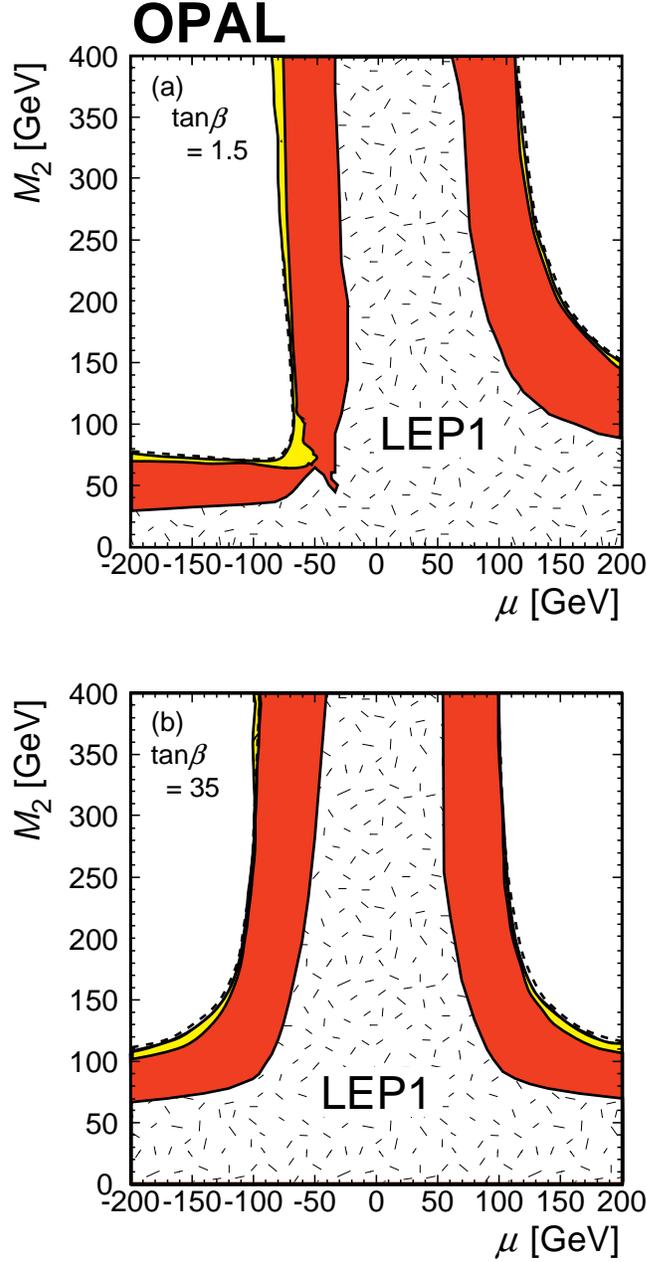, bbllx=150pt,bblly=174pt,bburx=393pt,bbury=655pt}
}\end{center}
\caption[1]
{
Exclusion regions at 95\% C.L. in the $\mu$-$M_2$ plane
within the framework of the Constrained MSSM
for (a) $\tan \beta=1.5$ and (b) $\tan \beta=35$.
The speckled areas show the regions excluded by
LEP1 data, and the shaded areas the additional
exclusion regions using the data from $\roots = 183$ and $189$~GeV.
The dark-shaded regions are valid
for all values of $m_0$, and the light-shaded
regions are the additional excluded parameters if
$m_0 \ge 500$~GeV (i.e., heavy scalar leptons).
The kinematical boundary at $\roots = 189$~GeV for
$\chp_1 \chm_1$ production is shown by
dashed lines.
}
\label{fig3}
\end{minipage}
\end{figure}
%-----------------------------------------------------------------------
\newpage
\begin{figure}
\begin{center}\mbox{
\epsfig{file=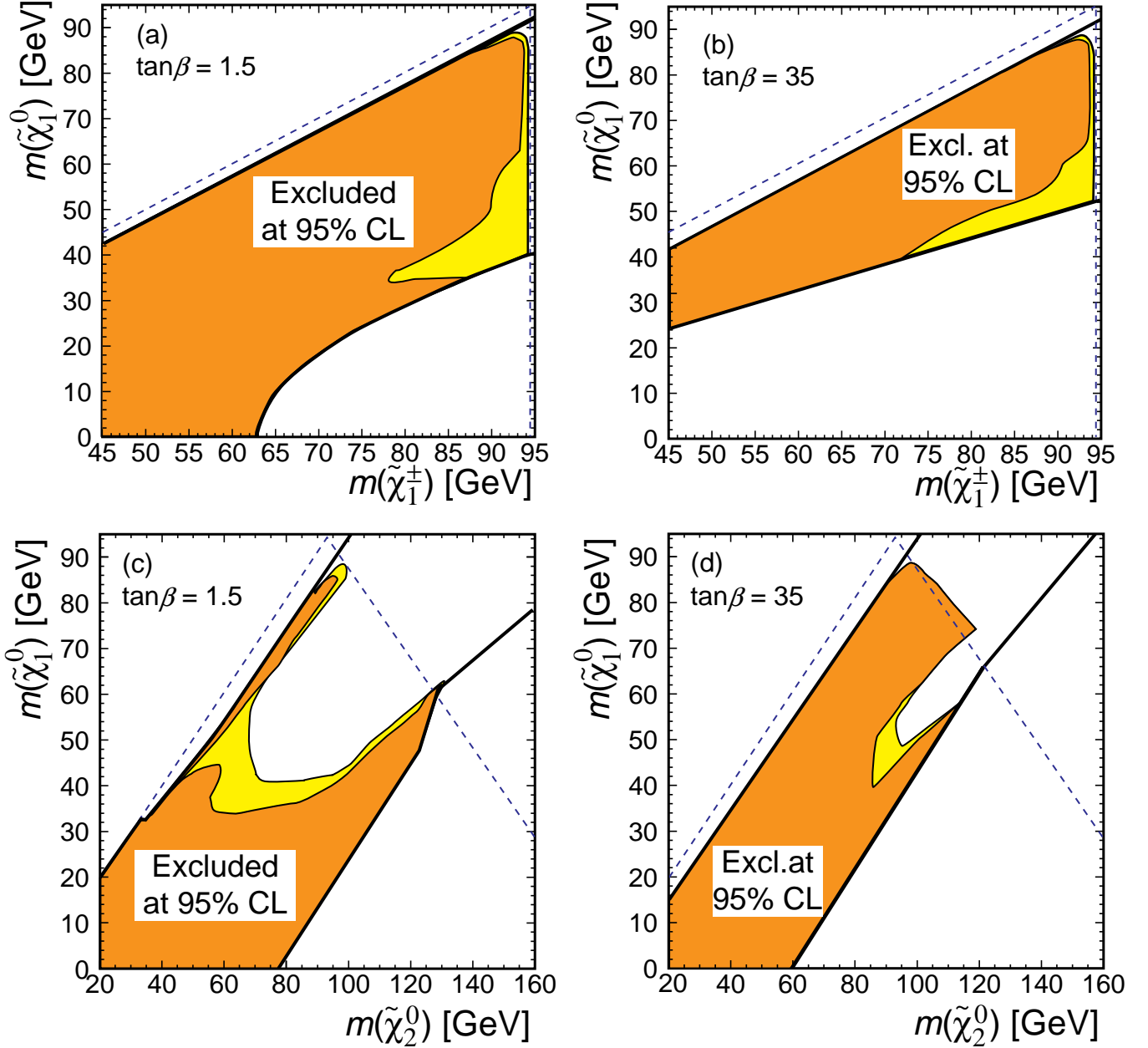,bbllx=28pt,bblly=199pt,bburx=538pt,bbury=701pt}
}\end{center}
\caption[1]
{
The 95\% C.L. excluded regions in the ($m_{\chp_1}$,$m_{\nt_1}$) plane
within the framework of the Constrained MSSM
for the case of any value of $m_0$ (dark-shaded regions) and
$m_0 \ge 500$~GeV (extending to light-shaded regions) for
(a) $\tan \beta=1.5$ and (b) $\tan \beta = 35$.
The thick solid lines represent the theoretical bounds
of the Constrained MSSM parameter space.
The kinematical boundaries for production and
decay at $\roots = 189$~GeV are shown by
dashed lines.
Similar plots for neutralino masses are shown
in (c) for $\tan \beta=1.5$ and (d) for $\tan \beta = 35$.
}
\label{fig4}
\end{figure}
%-----------------------------------------------------------------------
%-----------------------------------------------------------------------
\newpage
\begin{figure}
\begin{center}\mbox{
\epsfig{file=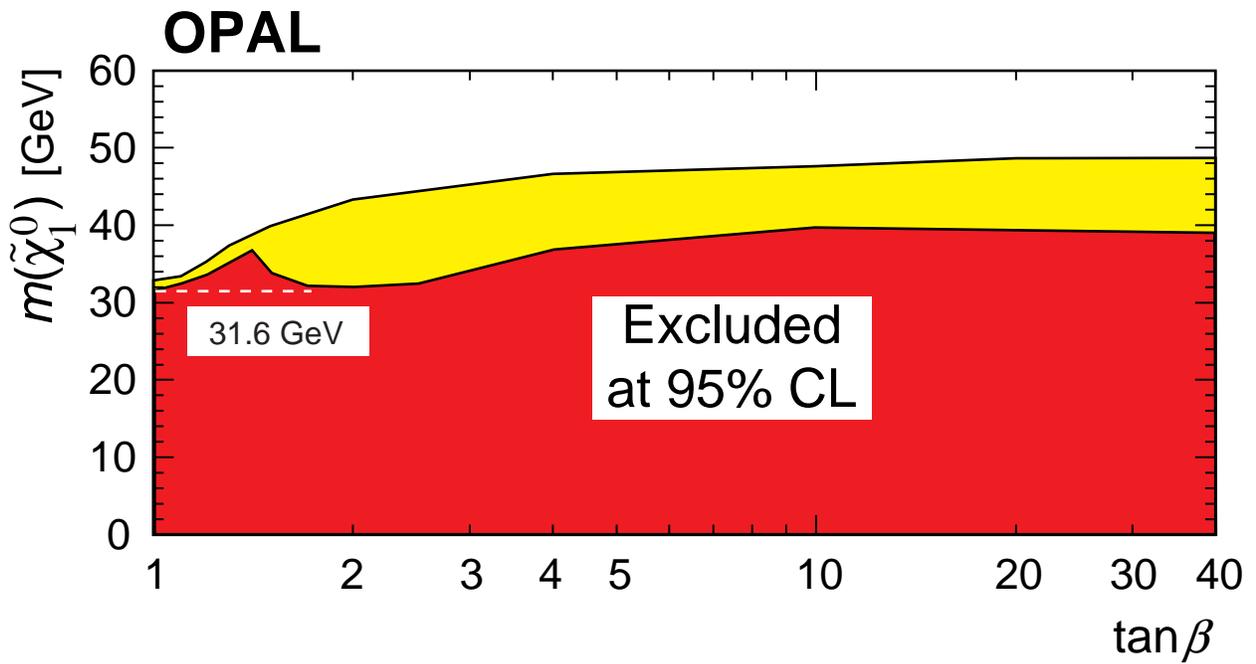,bbllx=72pt,bblly=390pt,bburx=537pt,bbury=647pt}
}\end{center}
\caption[1]
{
The 95\% C.L. mass limit on the lightest neutralino $\nt_1$ as
a function of $\tan\beta$ within the framework of the Constrained MSSM.
Exclusion regions for all $m_0$ values (dark-shaded region) and 
$m_0 \ge 500$~GeV (extending to the light-shaded region).}
\label{fig5}
\end{figure}
%-----------------------------------------------------------------------
%============================================================================
%%%%%%%%%%%%%%%%%%%%%%%%%%%%%%%%%%%%%%%%%%%%%%%%%%%%%%%%%%%%%%%%%%%%%%%%
\end{document}